# Magnetic structure evolution and magnetoelastic coupling across the spin reorientation transition in TmCrO$_3$

Vishesh Sharma[1], Gaurav Gautam[1], Poonam Yadav[2], Chin-Wei Wang[3], Kaya Wei[4], N. P. Lalla[5], Theo Siegrist[4,6] and Shivani Sharma[4*]

[1]*Department of Physics, Graphic Era Hill University, Dehradun, India 248002.*
[2]*Center for Integrated Nanostructure Physics, Institute for Basic Science, Suwon 16419, Republic of Korea.*
[3]*National Synchrotron Radiation Research Center, Hsinchu 300092, Taiwan*
[4]*National High Magnetic Field Laboratory, Tallahassee, FL 32310, United States.*
[5]*UGC-DAE Consortium for Scientific Research, Indore, MP India.*
[6]*Department of Chemical and Biomedical Engineering, FAMU-FSU College of Engineering, Florida State University, Tallahassee, FL 32310, United States*
[*]*Corresponding author: phy.shivanisharma@gmail.com*

## Abstract

We present a comprehensive study of the magnetic structure evolution across the spin reorientation transition in orthorhombic (*Pnma*) TmCrO$_3$. Magnetic susceptibility reveals canted antiferromagnetic (CAFM) ordering at $T_N$ = 125 K, two compensation points ($T_{comp1}$ and $T_{comp1}$), followed by the magnetization reversal with magnetic susceptibility minimum between $T_{comp1}$ and $T_{comp1}$. Heat-capacity shows a λ-transition at $T_N$, associated with long-range antiferromagnetic order of Cr, followed by a broad feature near 9 K. Neutron powder diffraction (NPD) establishes the *Pn'm'a*: Γ2($C_x, G_y; C_x^{Tm}, F_z^{Tm}$) magnetic structure below $T_N$. A gradual change in magnetic structure occurs during the spin-reorientation (SRO) transition below 30 K where the magnetic structure transitions from *Pn'm'a* (Γ2) to *Pn'ma'* (Γ4): $A_x, F_y, G_z; F_y^{Tm}$ phase. However, below SRO, neither Γ2 nor Γ4 alone adequately fit the intensity of magnetic reflections. A satisfactory refinement could be achieved in the monoclinic subgroup *P2$_1$'/c'* derived from a combination of Γ2 and Γ4. The gradual SRO of Tm and Cr moments across the compensation regime is consistent with the magnetic symmetry *P2$_1$'/c'*: $C_x, G_y, G_z; C_x^{Tm}, F_y^{Tm}, F_z^{Tm}$. Furthermore, the ordered moments Cr and Tm in TmCrO$_3$ exhibit a complex, nonmonotonic temperature dependence, with Tm driving a spin-reorientation transition near the compensation point. Anomalies in lattice parameters reveal strong magnetoelastic coupling, linking structural distortions to the SRO.

## Introduction

Rare earth (RE) perovskite oxides with transition metal (TM) ion (Fe, Cr, V) at the B-site have been extensively studied for exhibiting complex magnetism, including spin-reorientation

(SRO) transition, multiferroicity, spin-phonon coupling and temperature-induced magnetization reversal [1–5]. Most of these compounds crystallize in a centrosymmetric orthorhombic phase *Pnma* (No. 62, GdFeO$_3$-type ) where transition metal (TM) ions occupy the 4b site, forming a distorted TM-O$_6$ octahedra. With changing temperature, these distortions affect the magnetic response, leading to SRO transition, followed by magnetization reversal [6–8]. In these perovskites, magnetization reversal occurs during field-cooled cooling (FCC) curve below the TM-CAFM ordering temperature $T_N$. During FCC the net magnetization of the antiparallel polarized paramagnetic RE moments, against the net sublattice magnetization arising from the uncompensated TM moments of the CAFM order, overcompensates it below $T_{comp}$ [9–12]. Antiparallel polarization of the paramagnetic RE moments results from the negative super-exchange interaction between TM-O-RE. This phenomenon has been widely reported in rare-earth vanadates, ferrites, and chromates, which is supposed to arise from the delicate interplay between the 3d transition-metal and the 4f rare-earth sublattices [4,7,13].

Magnetization reversal is intriguing both for its fundamental importance and for applications in spintronic devices such as magnetic switches and spin-valves, which exploit two stable magnetic states under an external field [14,15]. In systems exhibiting this behavior, the SRO is typically followed by a compensation temperature $T_{comp}$, at which the sublattice magnetizations cancel each other, yielding zero net moment. Upon further cooling below $T_{comp}$, the imbalance between antiferromagnetically coupled sublattices drives the net magnetization negative, before recovering a positive value at lower temperatures. This trend of magnetic phenomena is observed for most of the orthochromates and orthoferrites [3,9,16–21]. On the other hand, in nonmagnetic rare-earth vanadate such as YVO$_3$, the magnetization reversal originates from the coexistence of competing magnetic phases, which is intimately coupled to the underlying structural phase transition [7]. The phenomenon of magnetization reversal has not only been observed in perovskites but has been widely observed in different classes of materials such as spinel, alloys, and garnets [22–27].

Bulk magnetization measurements on TmCrO$_3$ reveal long-range CAFM ordering at ~125 K, followed by two successive magnetization reversals: the first near 30 K and a second at lower temperatures around 11 K [9,28]. The first compensation temperature (~30 K) has previously been attributed to antiparallel coupling between the rare-earth (Tm) and Cr sublattices [29], whereas the origin of the second reversal in isostructural GdCrO$_3$ has recently been attributed to the flipping of net magnetization along the applied field (H) in order to minimize the Zeman energy [12]. In addition, TmCrO$_3$ exhibits both magnetocaloric and exchange bias effects [9,28]. Despite these observations, direct evidence for the evolution of the magnetic structure of TmCrO$_3$ across the compensation points is lacking. In this work, we present neutron powder diffraction, magnetization and heat capacity results that reveal the microscopic origin of these anomalies. Our analysis confirm that TmCrO$_3$ undergoes simultaneous AFM ordering of Tm and Cr ions at $T_N$ = 125 K. Below 50 K, the system enters a spin-reorientation regime involving a transition from the Γ2 and Γ4 configuration. However, instead of stabilizing in a single configuration, TmCrO$_3$ retains a mixed Γ2/ Γ4 state even down to 3.5 K, giving rise to a complex magnetic ground state with $P2_1'/c'$ monoclinic symmetry. Detailed analysis confirms the first-order spin reorientation transition, which is mainly governed by the Tm sublattice. These results provide direct evidence that the low-temperature magnetization reversals are closely tied to the coexistence of the reorientations of Tm and Cr spins. Additionally, clear anomalies are observed in the lattice parameters across the compensation points, highlighting strong magnetoelastic coupling in this system. These results confirm that the magnetization reversals in TmCrO$_3$ originate from the cooperative interaction between spin reorientation and lattice distortions.

**Experimental Section**

The solid-state reaction method has been used to synthesize polycrystalline TmCrO$_3$ [9]. Elemental composition of Thulium (Tm) and Chromium (Cr) in a phase-pure sample was determined using energy-dispersive X-ray spectroscopy (EDS) (FEI Quanta). DC Magnetization was measured using a Quantum Design SQUID magnetometer under zero-field cooled (ZFC) and field-cooled (FC) and field-cooled warming (FCW) conditions in an applied field of 0.1 T. Heat capacity was measured using a Quantum Design physical property measurement system (PPMS). Powder neutron diffraction measurements on TmCrO$_3$ were carried out at ECHIDNA and WOMBAT beamlines at the OPAL facility, ANSTO, Australia. Data from the WOMBAT beamline with a neutron wavelength of 2.41 Å, was used for the magnetic structure analysis in a temperatures range from 3.5 K to 120 K. The JANA2006 program suite was used for the magnetic structure refinement [30].

**Results and Discussion**

 [31]The same sample previously studied [31] is used for magnetic structure determination. Figure 1(a) shows the neutron Rietveld 200 K pattern of TmCrO$_3$ measured at WOMBAT with λ = 2.41 Å, that defines the paramagnetic background above the AFM ordering. All structural Bragg peaks are fitted with orthorhombic symmetry and space group *Pnma*. The results are given in Table 1, and the crystal structure is shown in Figure 1(b), where the distorted CrO$_6$ octahedra are shown in pink, Tm atoms in blue, and oxygen atoms in cyan.

Table 1: Structural parameters of TmCrO$_3$ at 200 K, determined using neutron data. a = 5.4931(4), b = 7.4891(8) and c = 5.2023(5) Å. The values of Goodness-of-fitting (GOF) and R$_{wp}$ factor are 2.01 and 5.97, respectively.

| Atoms | Wyckoff positions | | | Occupancies | Uiso (Å$^2$) |
|---|---|---|---|---|---|
| | x | y | z | | |
| Tm | 0.0717(8) | 0.25 | -0.018(1) | 1 | 0.029(2) |
| Cr | 0 | 0 | 0.5 | 1 | 0.020(2) |
| O1 | 0.460(1) | 0.25 | 0.106(1) | 1 | 0.030(2) |
| O2 | 0.3030(8) | 0.0556(5) | -0.3079(8) | 1 | 0.029(2) |

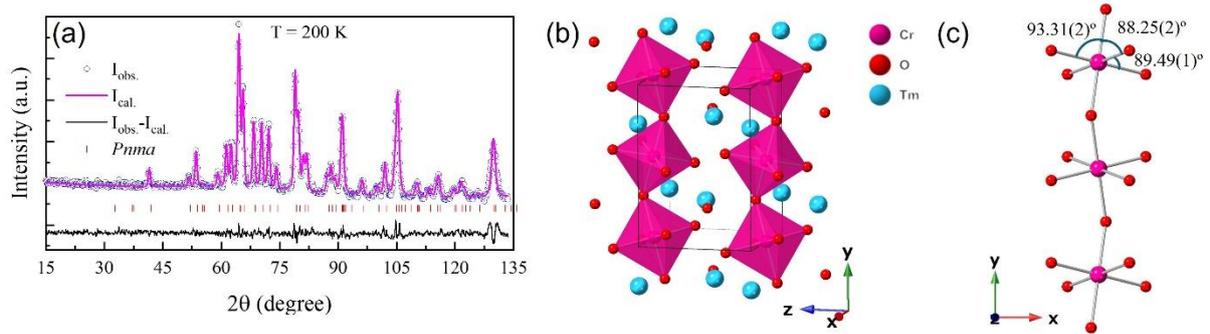

Figure 1: (a) Rietveld refined 200 K NPD pattern of TmCrO$_3$. (b) Crystal structure of TmCrO$_3$ made based on the refined parameters. This highlights the tilted CrO$_6$ octahedra (pink) forming a corner-shared network. (c) Octahedral chain along the *c*-axis highlighting the Cr–O–Cr bond angle and illustrating the distortions in the CrO$_6$ octahedra.

**Magnetization and heat capacity studies**

Figure 2(a) displays the zero-field cooled (ZFC), field cooled (FC) and field-cooled warming (FCW/FC) magnetic susceptibility ($\chi$) of TCO measured with a 0.1 T field. With decreasing temperature, a sharp upturn is observed below 126 K, followed by a peak at 123 K in the ZFC curve. The transition temperature is determined by the first order derivation of ZFC magnetic susceptibility which gives $T_N$ = 125 K, consistent with literature [9,32]. The inset shows the enlarged view around $T_N$. Below $T_N$, with decreasing temperature, the FC susceptibility increases with decreasing temperature until 60 K, below which it starts decreasing continuously, leading to a crossover between the ZFC and FC curves. Below that, the magnetization turns negative after crossing the y-axis at the compensation temperatures. In the present system, two such compensation points are identified, denoted as $T_{comp1}$ and $T_{comp2}$. As shown in Fig. 2(a), these correspond to the intersections of the magnetization curve with the zero-magnetization line (horizontal dashed line) at approximately 18 K and 8 K. Notably, the magnetization reaches its most negative value at 11 K, highlighting the strong antiparallel alignment of the sublattices in this regime. Below 11 K, the susceptibility increases before saturating at lower temperatures. The field-cooled warming curve (FCW/FC) follows the ZFC curve. The first order derivative of $\chi$ is presented in Fig. 2(b). It shows two peaks; first at $T_N$ and then another at 11 K, below which the magnetization increases sharply and reverts to the positive value. The inset in Fig. 2(b) shows the inverse susceptibility (1/$\chi$) of ZFC curve measured with 1 T. The Curie-Weiss (CW) fit of 1/$\chi$ shows linear behaviour down to 132 K and yields a value for the Curie constant C = 9.2 corresponding to $\mu_{eff}$ = 8.58 $\mu_B$/f.u., close to the expected value of 8.48 $\mu_B$. The Weiss constant of $\Theta$ = -33 K indicates AFM interactions, consistent with the M-H isotherms [31].

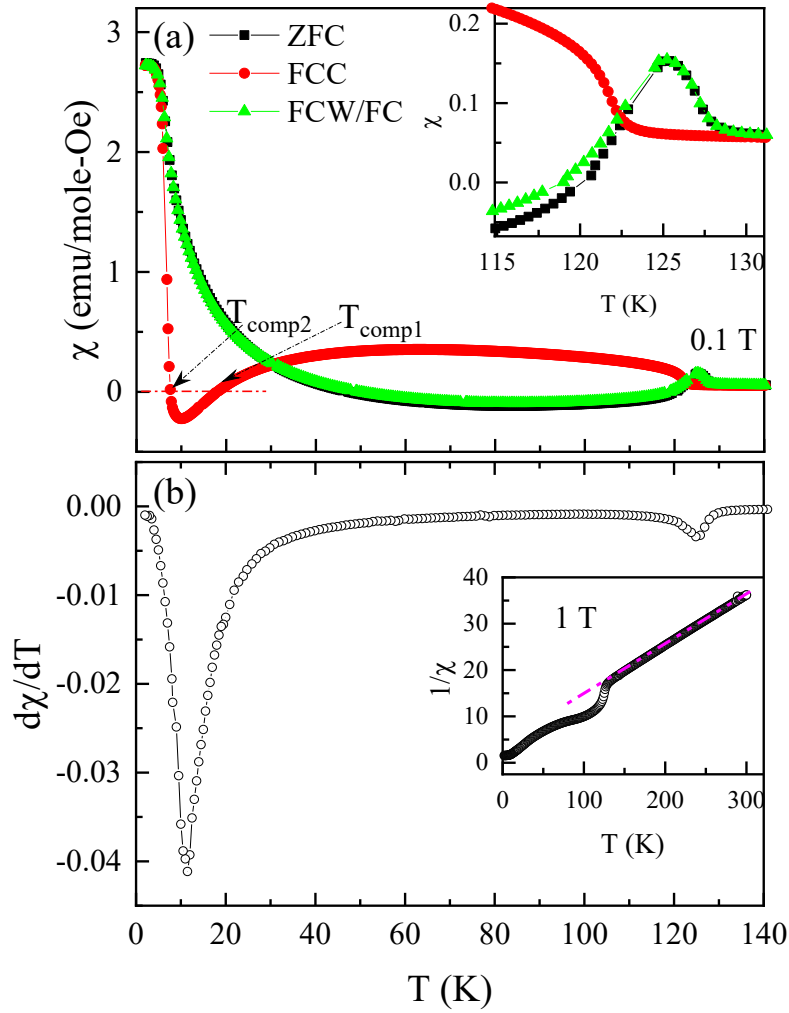

**Figure 2:** (a) Temperature-dependent DC magnetic susceptibility ($\chi_{dc}$) of TCO measured at B = 0.1 T field under ZFC, FCC and FCW conditions. The inset in panel (a) shows the enlarged view around the first transition, highlighted by the red box in the main figure. Arrows indicate the compensation temperatures $T_{comp1}$ and $T_{comp2}$. (b) First-order derivative of ZFC $\chi_{dc}$ (0.1 T) exhibits the first peaks at $T_N$ and between $T_{comp1}$ and $T_{comp2}$. The inset to panel (b) presents the Curie-Weiss fit of the inverse susceptibility measured under 1 T.

To further investigate the nature of these magnetic transitions, heat capacity has been measured between 1.8 and 200 K under zero magnetic field, shown in Figure 3. A λ-type transition is observed at 125 K, associated with the long-range CAFM order of $Cr^{3+}$. With decreasing temperature, another broad feature is observed around a temperature of 9 K. This broad peak likely originates from the crystal field levels of $Tm^{3+}$, similar to $YbCrO_3$ [18]. Schottky anomalies are commonly observed for various rare-earth-based systems [18,33–35]. However, in single crystalline $TmCrO_3$, in addition to $T_N$, another transition was observed at 19.6 K which was attributed to the Tm ordering at that temperature. However, consistent with our results, this transition was not reported for polycrystalline $TmCrO_3$ [9].

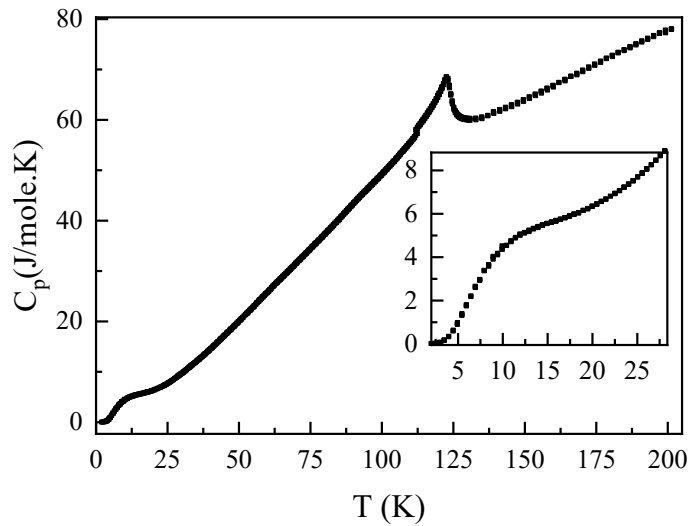

**Figure 3**: Temperature-dependent heat capacity of TmCrO$_3$. The inset shows the zoom view for the temperature range of 1.8 to 28 K.

**Neutron powder diffraction**

To investigate the magnetic structure as a function of temperature, NPD measurements have been performed at WOMBAT. Figure 4 shows the NPD patterns of TCO measured over a range of temperatures. Below $T_N$, several additional peaks emerge, with the two most intense magnetic reflections, (110) and (011), appearing near $2\theta = 30°$ as shown in Figure 4(a). The waterfall NPD plots as a function of temperature are given in Figure S1. Figures 4(b) and 4(c) depict the temperature-dependent peak profiles and integrated intensities of the (110) and (011) peaks below $T_N$. The intensities of both peaks increase upon cooling down to 30 K, below which they exhibit distinct temperature dependencies. Below it, the intensity of the (110) reflection increases sharply, while that of the (011) reflection decreases. This divergence suggests a reorientation of the magnetic moments of Tm$^{3+}$ and/or Cr$^{3+}$, possibly reflecting changes in the relative contributions of the Tm and Cr sublattices to the net magnetic structure. The behavior is linked to the compensation effect, where sublattice magnetizations balance and then reverse dominance as the temperature decreases. In Fig. 4(c), the blue lines serve as a guide to the eye, showing that below 30 K the temperature dependence of (110) and (011) differs markedly. The intensity of (110) increases below 30 K, while for (011), it decreases.

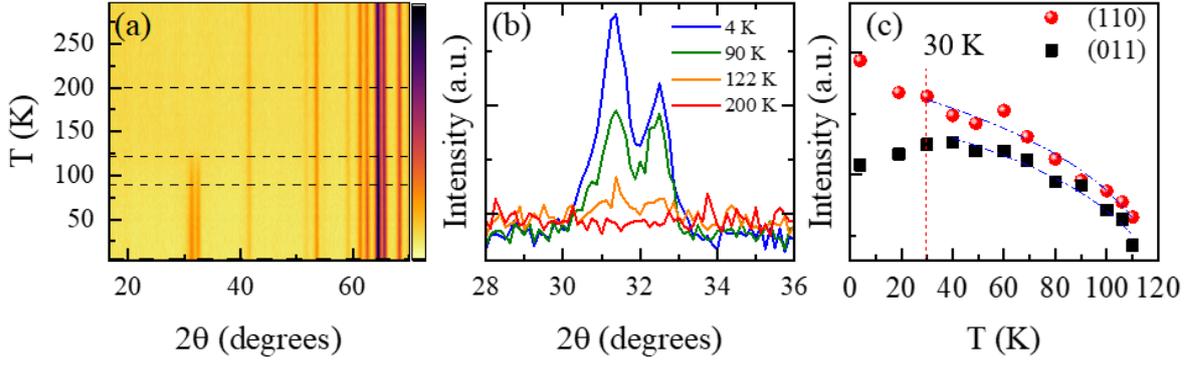

**Figure 4:** (a) Thermo-neutron diffractogram of TmCrO$_3$, showing the evolution of magnetic Bragg reflections below $T_N$. (b) Temperature-dependent intensity profiles of the (110) and (011) magnetic reflections. (c) Integrated intensity of the (110) and (011) reflections as a function of temperature. The vertical dashed line marks the boundary below which the peak shows different temperature dependence. Blue dashed lines serve as a guide to the eye.

All the magnetic reflections observed below $T_N$ can be indexed by the propagation vector **k** = (0, 0, 0). For **k** = (0, 0, 0), with Tm and Cr being the magnetic ions, the group analysis results in eight possible maximal magnetic space-groups for the parent space-group *Pnma*. Out of the eight possible magnetic space groups, only four, *Pnma* (Γ1), *Pn'm'a* (Γ2), *Pnm'a'* (Γ3), and *Pn'ma'* (Γ4) yield finite moments at the Cr site. The remaining four models were therefore discarded. The allowed magnetic configurations Γ1 and Γ3 failed to match the observed intensity of magnetic peaks, and can be ruled out. The remaining models, Γ2 and Γ4, were further tested against the data below $T_N$, and the corresponding fits are shown in figure 5. The fit quality of both the models was evaluated by examining the most prominent magnetic peaks (1 1 0) and (0 1 1), [see Figs. 5(a, b)], which clearly shows improved agreement with the observed data in the case of Γ2, especially down to 25 K. The full patterns at 120 K and 50 K, fitted with Γ2, are given in the supplementary information as Figure S2 [36]. For refining the magnetic structure in Γ2 configuration, the symmetry permits relaxation of all the three magnetic moment components (Ma, Mb, and Mc) for the Cr ions. However, the fitting with all the three components relaxed, results in nearly negligible values for Ma and Mc. To test this further, we refined the data by constraining Mc = 0, and separately by constraining both Ma and Mc to 0. Setting Mc = 0 does not alter the refinement quality, as the reliability factors remain unchanged compared to that of the unconstrained model. In contrast, fixing both Ma and Mc to zero leads to a slight decrease in the fit quality, indicating a non-zero Ma. For Tm, the Ma and Mc components are symmetry-allowed and were refined under the constraint Ma = −Mc. Following Shamir's notation [37], the magnetic structure in the Γ2 phase can be described as $Pn'm'a: \Gamma2(C_x, G_y; C_x^R, F_z^R)$, where the magnetic structure of Cr ions are C-type along *x* and G type along *y* axis whereas the rare-earth Tm moments have C-type ordering along *x* and F-type along *z* directions. Similar notations were recently used to define magnetic structures for other members of this family [13,17].

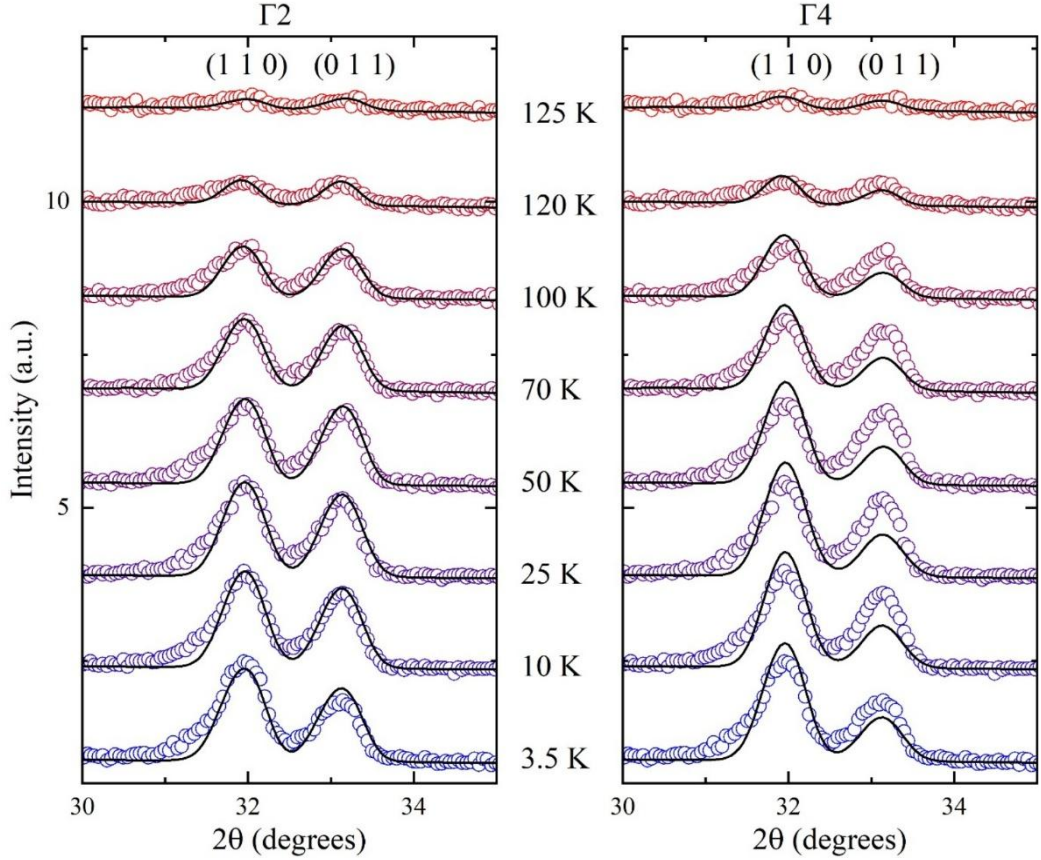

Figure 5: The zoomed-in view depicting the refinement quality of the most prominent magnetic peaks (1 1 0) and (0 1 1) using magnetic space groups (a) *Pn'm'a* (Γ2) and (b) *Pn'ma'* (Γ4).

For 25 K < T < $T_N$, the *Pn'm'a* (Γ2) magnetic model fits the data very well. However, below 25 K, discrepancies begin to appear—specifically, the (1 1 0) peak remains underfitted, while the (0 1 1) peak gets overfitted. In contrast, the *Pn'ma'* (Γ4) model provides a poor fit at 25 K and above, though it's fitting improves slightly below 25 K. For example, at 3.5 K, with Γ4, the (1 1 0) peak is overfitted, and the (0 1 1) peak is underfitted, which is opposite to Γ2 model. Thus, neither of the models, while taken individually, provide satisfactory fit at 3.5 K. A close review of the diffraction data suggests that below 25 K, the transition involves a gradual transformation of *Pn'm'a* (Γ2) phase to *Pn'ma'* (Γ4) phase. Therefore, at 10 K and 3.5 K, the system remains in an intermediate state, with spins attaining orientations intermediate to these two magnetic structures. Also, the gradual change of spin orientations from (Γ2) to (Γ4) demands that thermodynamically the phase-transition must be first-order, which allows coexistence of both the phases. Moreover, the magnetization data indicate an SRO below $T_{comp1}$, which is unambiguously supported by the diffraction results.

At 10 K and 3.5 K, the TCO appears to be in an intermediate state. To fit the data at these temperatures, the magnetic symmetry was reduced to $P2_1'/c'$, a subgroup systematically derived from both *Pn'm'a* (Γ2) and *Pn'ma'* (Γ4). This subgroup was identified using the K-SUBGROUPSMAG utility of the Bilbao Crystallographic Server, which is given in Figure 6 [38].

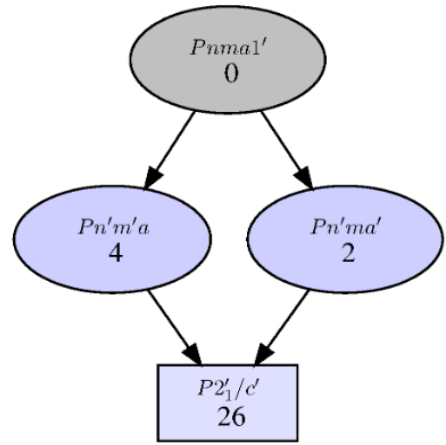

**Figure 6**: Diagram illustrating the subgroup lowered from the Γ2 and Γ4 magnetic space groups (MSG), generated using the K-SUBGROUPSMAG tool on the Bilbao Crystallographic Server [38].

Figures 7(a-c) presents comparative fitting of the most prominent magnetic reflections using different models, *Pn'm'a* (Γ2), *Pn'ma'* (Γ4), and $P2_1'/c'$. Among these, $P2_1'/c'$ fits the data best, as indicated by better quality fit along with the lowest $R_p$ and $R_{wp}$ values. In the monoclinic $P2_1'/c'$ symmetry, the Cr site splits into two positions which was constrained to have identical moments with opposite signs for all components. All the Mx, My and Mz components for both Tm and Cr were allowed to be refined. However, refining all of them without the prior knowledge of magnetic structure above $T_{comp1.}$, was a non-trivial task. Therefore, after several iterations, the best quality of fit was achieved with refining the z component of Cr fixed to 0 in Γ2. Also, the *y* component of Tm was refined in addition to *x* and *z* component with the restriction; Mx = -Mz. For Cr1 and Cr2, the sign of *x*, *y* and *z* components were opposite. The final magnetic structure below SRO transition can be best described as $P2_1'/c'$ ($C_x, G_y, G_z;\ C_x^z, F_y^z, F_z^z$). The resulting magnetic structures corresponding to different models are illustrated in Fig. 7. The allowed magnetic moment components in different magnetic symmetries are summarized in table 2.

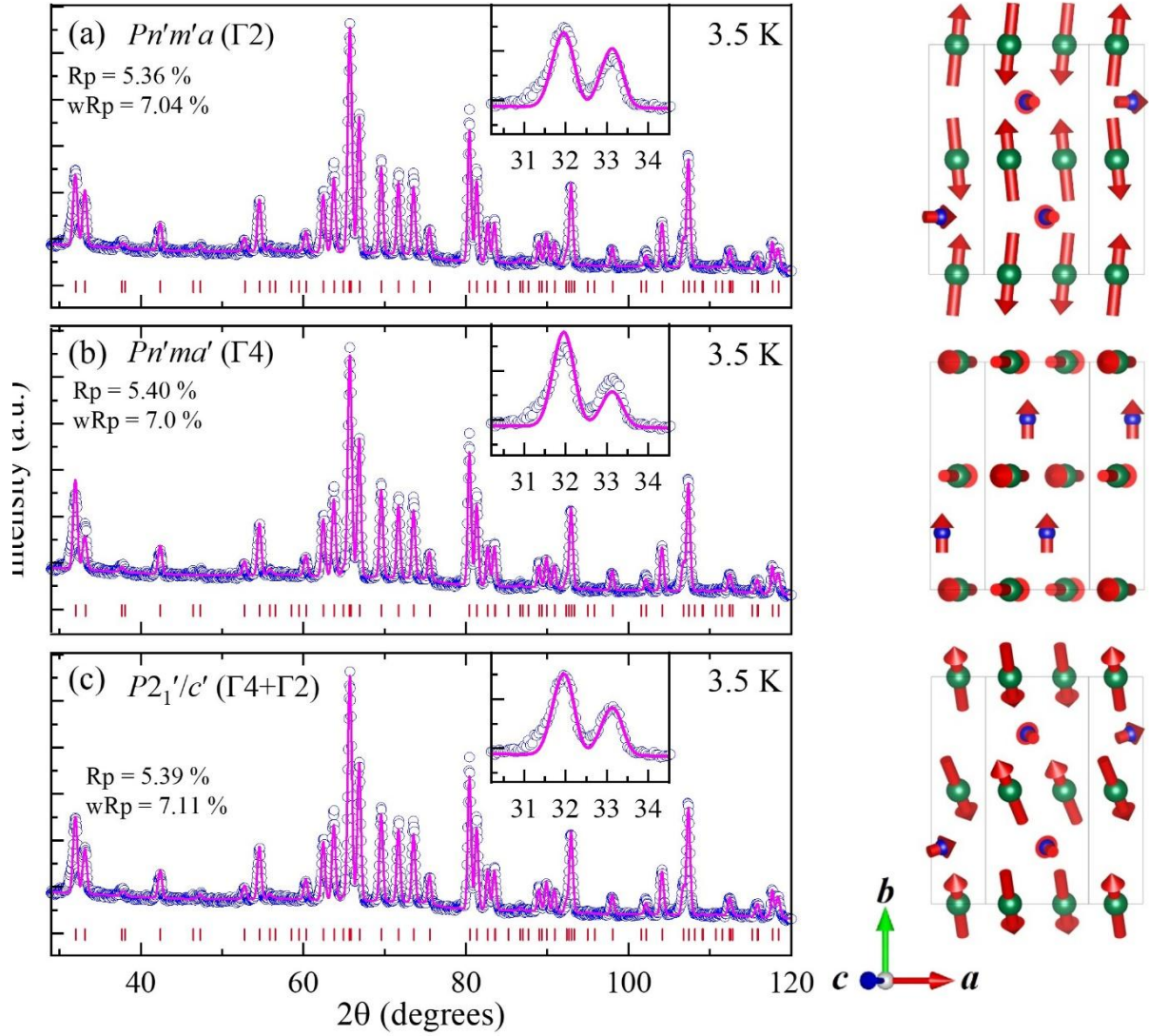

**Figure 7:** Comparison of the observed and calculated magnetic diffraction profiles at 3.5 K for the three tested magnetic models: $Pn'm'a$ ($\Gamma2$), $Pn'ma'$ ($\Gamma4$), and $P2_1'/c'$. Open circles represent the experimental data, and solid lines correspond to the calculated patterns from Rietveld refinement. The reliability factors (Rp and R$_{wp}$) are indicated for each refinement. The corresponding magnetic structures for Cr (green) and Tm (blue) moments are shown on the right, with arrows indicating the direction of the magnetic moments along the crystallographic axes.

**Table 2**: Magnetic space group for each magnetic structure models for TmCrO$_3$.

| Magnetic space group | Irrep. | Cr | | | Tm | | |
|---|---|---|---|---|---|---|---|
| $Pn'm'a$ | $\Gamma2$ | $C_x$ | $G_y$ | $F_z$ (=0) | $C_x^{Tm}$ | | $F_z^{Tm}$ |
| $Pn'ma'$ | $\Gamma4$ | $A_x$ | $F_y$ | $G_z$ | | $F_y^{Tm}$ | |
| $P2_1'/c'$ | $\Gamma2+\Gamma4$ | $C_x$ | $G_y$ | $G_z$ | $C_x^{Tm}$ | $F_y^{Tm}$ | $F_z^{Tm}$ |

The temperature dependence of the ordered moments of Tm ($M_{Tm}$) and Cr ($M_{Cr}$), as obtained from refinement, is shown in Fig. 8(a). Upon cooling, the $M_{Cr}$ moment increases rapidly below $T_N$ and saturates at ~2.1 $\mu_B$ around 50 K, and remains nearly constant thereafter. In contrast, $M_{Tm}$ exhibits a nonmonotonic temperature evolution. It decreases slightly from 0.37 $\mu_B$ to 0.17 $\mu_B$ near 70 K, then increases again with a sharp enhancement appearing only below $T_{comp}$. To highlight the direct correlation with the bulk magnetic response, the ZFC magnetization is overlaid in the background (purple). The anomalous variation of $M_{Tm}$ mirrors the nontrivial behavior of the dc susceptibility: $\chi_{dc}$ initially increases below $T_N$, peaks near 60 K, and then decreases, eventually dropping below its value above $T_N$. Furthermore, the sharp rise in $M_{Tm}$ below $T_{comp}$ coincides with the marked decrease in magnetization, while the additional increase in magnetization below 10 K corresponds to a subtle reduction of $M_{Cr}$ at 3.5 K. This one-to-one correspondence between the evolution of the ordered moments and the bulk magnetization highlights a complex spin-reorientation process in TCO, where both Cr and Tm moments participate. However, the pronounced temperature dependence of $M_{Tm}$ clearly indicates that the Tm sublattice plays the dominant role in driving the SRO.

To further explore the structural link of the SRO transition, the temperature evolution of the lattice parameters is presented in Fig. 8(b). All three lattice constants decrease monotonically upon cooling down to ~ 60 K. Below this temperature, the $c$ parameter exhibits an anomalous increase down to $T_{comp1}$, followed by a significant decrease below $T_{comp2}$. The abrupt increase of lattice parameter $c$ for temperatures from 60 to 30 K, clearly indicates negative thermal expansion. In contrast, the $a$ and $b$ parameters continue to contract across the entire range, albeit with distinct slopes above and below compensation points. Earlier similar non-monotonic changes in lattice parameters were reported for $YbCrO_3$ which were explained based on temperature variations in Cr-Cr distances via O1 and O2 atoms [18]. It was observed that along O2, the Cr-Cr distance increases, while via O1, it decreases, which led to such non-monotonic variations in lattice parameters. Similar effects were observed in [39]. In the present case, the lattice parameter $c$ increases approximately by 0.02 % from 60 to 30 K, which is consistent with the reports on other such systems. The inset of Fig. 8(b) shows the temperature dependence of the lattice volume. While the overall volume decreases continuously with decreasing temperature, the rate of contraction differs across temperature regimes. Above $T_{comp1}$, the decrease is gradual, whereas below $T_{comp1}$, it becomes significantly more pronounced. These anomalies demonstrate a clear coupling between the lattice and the magnetic response of $TmCrO_3$, highlighting the role of structural distortions in stabilizing the spin reorientation.

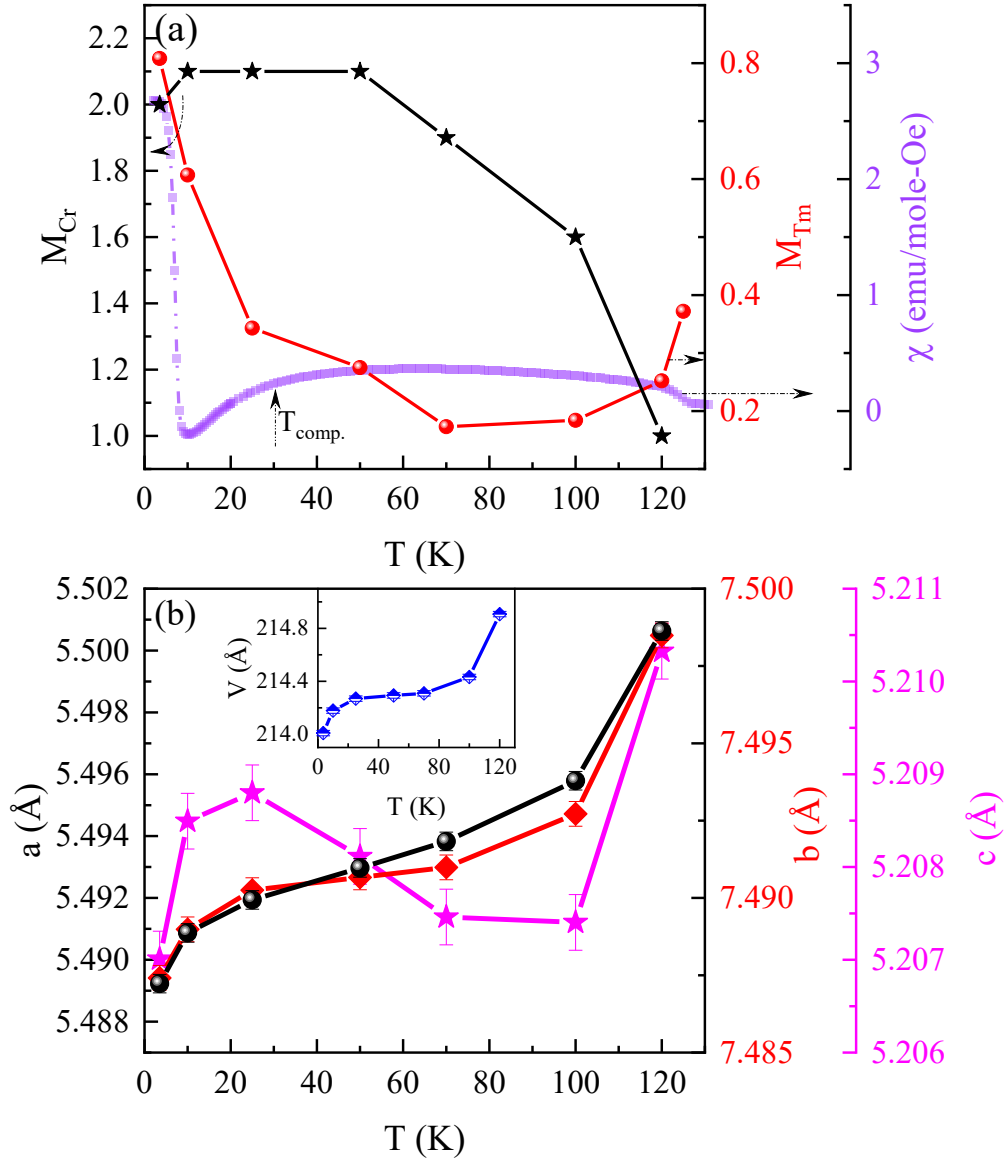

Figure 8. (a) Temperature dependence of the ordered moments of Cr (black) and Tm (red), plotted together with the ZFC susceptibility (purple, right axis). (b) Temperature evolution of the lattice parameters. The inset to (b) shows the variation in volume expansion as a function of temperature.

**Conclusion**

In summary, we have investigated the structural and magnetic properties of TmCrO$_3$ using neutron powder diffraction, dc magnetometry, and heat capacity measurements. Magnetic susceptibility and heat capacity data reveal long-range antiferromagnetic ordering at $T_N \approx 125$ K, a first compensation point at 30 K, and second compensation below 11 K, with Curie–Weiss analysis indicating dominant AFM interactions. Below $T_N$, neutron diffraction identifies the Γ2 (*Pn'm'a*) magnetic structure below $T_N$. Upon further cooling, a complex spin-reorientation transition occurs, driven predominantly by the Tm sublattice, which gradually evolves toward the Γ4 (*Pn'ma'*) structure. At the lowest temperatures, refinements in the intermediate $P2_1'/c'$ symmetry provide the best description of the magnetic structure, capturing the nonmonotonic evolution of Cr and Tm moments. The observed anomalies in lattice parameters across the spin-reorientation transition demonstrate significant magnetoelastic coupling, highlighting the interplay between structural distortions and magnetic ordering. These results establish $TmCrO_3$ as a prototypical system exhibiting coupled spin-lattice dynamics and compensation-driven spin reorientation.

**Acknowledgement**

A portion of this work was performed at the National High Magnetic Field Laboratory, which is supported by National Science Foundation Cooperative Agreement No. DMR-2128556 and the State of Florida.

References:

[1] T. Kimura, G. Lawes, T. Goto, Y. Tokura, and A. P. Ramirez, Magnetoelectric phase diagrams of orthorhombic $R{\mathrm{MnO}}_{3}$ ($R=\mathrm{Gd}$, Tb, and Dy), Phys Rev B **71**, 224425 (2005).

[2] T. Kimura, T. Goto, H. Shintani, K. Ishizaka, T. Arima, and Y. Tokura, Magnetic control of ferroelectric polarization, Nature **426**, 55 (2003).

[3] S. Mahana, B. Rakshit, R. Basu, S. Dhara, B. Joseph, U. Manju, S. D. Mahanti, and D. Topwal, Local inversion symmetry breaking and spin-phonon coupling in the perovskite ${\mathrm{GdCrO}}_{3}$, Phys Rev B **96**, 104106 (2017).

[4] T. Yamaguchi and K. Tsushima, Magnetic Symmetry of Rare-Earth Orthochromites and Orthoferrites, Phys Rev B **8**, 5187 (1973).

[5] Y. Tokunaga, S. Iguchi, T. Arima, and Y. Tokura, Magnetic-Field-Induced Ferroelectric State in ${\mathrm{DyFeO}}_{3}$, Phys Rev Lett **101**, 97205 (2008).

[6] J. Hemberger, S. Lobina, H.-A. Krug von Nidda, N. Tristan, V. Yu. Ivanov, A. A. Mukhin, A. M. Balbashov, and A. Loidl, Complex interplay of $3d$ and $4f$ magnetism in ${\mathrm{La}}_{1\ensuremath{-}x}{\mathrm{Gd}}_{x}\mathrm{Mn}{\mathrm{O}}_{3}$, Phys Rev B **70**, 24414 (2004).

[7] S. Sharma, P. N. Shanbhag, F. Orlandi, P. Manuel, S. Langridge, D. Adroja, M. K. Sanyal, and A. Sundaresan, Symmetry Origin of the Dzyaloshinskii–Moriya Interaction and Magnetization Reversal in YVO3, Inorg Chem **60**, 2195 (2021).

[8] A. Kumar and S. M. Yusuf, The phenomenon of negative magnetization and its implications, Phys Rep **556**, 1 (2015).

[9] K. Yoshii, Magnetization reversal in TmCrO3, Mater Res Bull **47**, 3243 (2012).

[10] K. Yoshii, Magnetic Properties of Perovskite GdCrO3, J Solid State Chem **159**, 204 (2001).

[11] L. H. Yin, J. Yang, X. C. Kan, W. H. Song, J. M. Dai, and Y. P. Sun, Giant magnetocaloric effect and temperature induced magnetization jump in GdCrO3 single crystal, J Appl Phys **117**, 133901 (2015).

[12] P. Jain, S. Sharma, R. Baumbach, A. K. Yogi, I. Ishant, M. Majumder, T. Siegrist, M. K. Chattopadhyay, and N. P. Lalla, Structural role in temperature-induced magnetization reversal revealed in distorted perovskite $\mathrm{G}{\mathrm{d}}_{1--x}{\mathrm{Y}}_{x}\mathrm{Cr}{\mathrm{O}}_{3}$, Phys Rev B **109**, 94410 (2024).

[13] T. Sau, P. Yadav, S. Sharma, R. Raghunathan, P. Manuel, V. Petricek, U. P. Deshpande, and N. P. Lalla, High-resolution time of flight neutron diffraction and

magnetization studies of spin reorientation and polar transitions in ${\mathrm{SmCrO}}_{3}$, Phys Rev B **103**, 144418 (2021).

[14] G. Finocchio, I. N. Krivorotov, L. Torres, R. A. Buhrman, D. C. Ralph, and B. Azzerboni, Magnetization reversal driven by spin-polarized current in exchange-biased nanoscale spin valves, Phys Rev B **76**, 174408 (2007).

[15] A. Hirohata, K. Yamada, Y. Nakatani, I.-L. Prejbeanu, B. Diény, P. Pirro, and B. Hillebrands, Review on spintronics: Principles and device applications, J Magn Magn Mater **509**, 166711 (2020).

[16] T. Sau, S. Sharma, P. Yadav, R. Baumbach, T. Siegrist, A. Banerjee, and N. P. Lalla, First-order nature of the spin-reorientation phase transition in $\mathrm{SmCr}{\mathrm{O}}_{3}$, Phys Rev B **106**, 64413 (2022).

[17] S. Yano, C.-W. Wang, Y. Zhu, K. Sun, and H.-F. Li, Magnetic structure and phase transition in a single crystal of ${\mathrm{ErCrO}}_{3}$, Phys Rev B **108**, 174406 (2023).

[18] Deepak, A. Kumar, and S. M. Yusuf, Intertwined magnetization and exchange bias reversals across compensation temperature in $\mathrm{YbCr}{\mathrm{O}}_{3}$ compound, Phys Rev Mater **5**, 124402 (2021).

[19] R. D. Pierce, R. Wolfe, and L. G. Van Uitert, Spin Reorientation in Mixed Samarium-Dysprosium Orthoferrites, J Appl Phys **40**, 1241 (1969).

[20] J.-H. Lee, Y. K. Jeong, J. H. Park, M.-A. Oak, H. M. Jang, J. Y. Son, and J. F. Scott, Spin-Canting-Induced Improper Ferroelectricity and Spontaneous Magnetization Reversal in $\mathrm{SmFe}{\mathrm{O}}_{3}$, Phys Rev Lett **107**, 117201 (2011).

[21] C.-Y. Kuo et al., $k=0$ Magnetic Structure and Absence of Ferroelectricity in ${\mathrm{SmFeO}}_{3}$, Phys Rev Lett **113**, 217203 (2014).

[22] R. Padam, S. Pandya, S. Ravi, A. K. Nigam, S. Ramakrishnan, A. K. Grover, and D. Pal, Magnetic compensation effect and phase reversal of exchange bias field across compensation temperature in multiferroic Co(Cr0.95Fe0.05)2O4, Appl Phys Lett **102**, 112412 (2013).

[23] B. Meng, Q. S. Fu, X. H. Chen, G. S. Gong, C. Chakrabarti, Y. Q. Wang, and S. L. Yuan, Effect of Al substitution on the magnetization reversal and complex magnetic properties of NiCr2O4 ceramics, Physical Chemistry Chemical Physics **24**, 4925 (2022).

[24] M. K. Ray, B. Maji, K. Motla, S. K. P., and R. P. Singh, Multiple magnetization reversal and field induced orbital moment switching in intermetallic SmMnSi compound, J Appl Phys **128**, 073909 (2020).

[25] M. Oster, V. Ksenofontov, M. Dürl, and A. Möller, Giant Negative Magnetization in Al9–xFexMo3, Chemistry of Materials **31**, 9317 (2019).

[26] R. Pauthenet, Spontaneous Magnetization of Some Garnet Ferrites and the Aluminum Substituted Garnet Ferrites, J Appl Phys **29**, 253 (1958).

[27] M. Deb, P. Molho, B. Barbara, and J.-Y. Bigot, Controlling laser-induced magnetization reversal dynamics in a rare-earth iron garnet across the magnetization compensation point, Phys Rev B **97**, 134419 (2018).

[28] J. Shi, M. S. Seehra, J. Pfund, S. Yin, and M. Jain, Magnetocaloric properties of TbCrO3 and TmCrO3 and their comparison with those of the other RCrO3 systems (R = Gd, Dy, Ho, and Er), J Appl Phys **134**, 103903 (2023).

[29] R. M. Hornreich, Magnetic interactions and weak ferromagnetism in the rare-earth orthochromites, J Magn Magn Mater **7**, 280 (1978).

[30] V. Petříček, M. Dušek, and L. Palatinus, Crystallographic Computing System JANA2006: General features, Z Kristallogr Cryst Mater **229**, 345 (2014).

[31] V. Sharma, S. Pokhriyal, D. M. Stojr, A. Sharma, N. P. Lalla, and S. Sharma, Structure, Morphology and Magnetic Properties of TmCrO3, AIP Conference Proceeding (2026).

[32] J. A. Hodges, P. Imbert, and A. Schuhl, TmTO3 (T = Al, Fe, Cr and V). A 169Tm Mössbauer study of magnetic and crystal field properties, J Magn Magn Mater **43**, 101 (1984).

[33] S. J. Yuan, Y. M. Cao, L. Li, T. F. Qi, S. X. Cao, J. C. Zhang, L. E. DeLong, and G. Cao, First-order spin reorientation transition and specific-heat anomaly in CeFeO3, J Appl Phys **114**, 113909 (2013).

[34] X. F. Sun, I. Tsukada, T. Suzuki, S. Komiya, and Y. Ando, Large magnetothermal effect and spin-phonon coupling in a parent insulating cuprate ${\mathrm{Pr}}_{1.3}{\mathrm{La}}_{0.7}{\mathrm{CuO}}_{4}$, Phys Rev B **72**, 104501 (2005).

[35] Q. J. Li, Z. Y. Zhao, H. D. Zhou, W. P. Ke, X. M. Wang, C. Fan, X. G. Liu, L. M. Chen, X. Zhao, and X. F. Sun, Paramagnetic ground state with field-induced partial order in Nd${}_{3}$Ga${}_{5}$SiO${}_{14}$ probed by low-temperature heat transport, Phys Rev B **85**, 174438 (2012).

[36] V. Sharma, G. Gaurtam, P. Yadav, C.-W. Wang, K. Wei, N. P. Lalla, T. Siegrist, and S. Sharma, Supplementary Material, n.d.

[37] N. Shamir, H. Shaked, and S. Shtrikman, Magnetic structure of some rare-earth orthochromites, Phys Rev B **24**, 6642 (1981).

[38] M. I. Aroyo, J. M. Perez-Mato, C. Capillas, E. Kroumova, S. Ivantchev, G. Madariaga, A. Kirov, and H. Wondratschek, Bilbao Crystallographic Server: I. Databases and crystallographic computing programs:, Z Kristallogr Cryst Mater **221**, 15 (2006).

[39] F. Pomiro, R. D. Sánchez, G. Cuello, A. Maignan, C. Martin, and R. E. Carbonio, Spin reorientation, magnetization reversal, and negative thermal expansion observed in $R\mathrm{Fe}_{0.5}\mathrm{Cr}_{0.5}\mathrm{O}_{3}$ perovskites $(R=\mathrm{Lu},\mathrm{Yb},\mathrm{Tm})$, Phys Rev B **94**, 134402 (2016).

Supplemental Information

Fig. S1 presents temperature-dependent neutron powder diffraction (NPD) patterns of $TmCrO_3$, illustrating the evolution of magnetic reflections across the measured temperature range. As the temperature decreases, two diffraction peaks emerge in the low-angle region near $2\theta = 30°$, signifying the onset of long-range magnetic ordering. The distinct emergence and growth of magnetic Bragg peaks at lower temperatures confirm the establishment of a long-range magnetic order.

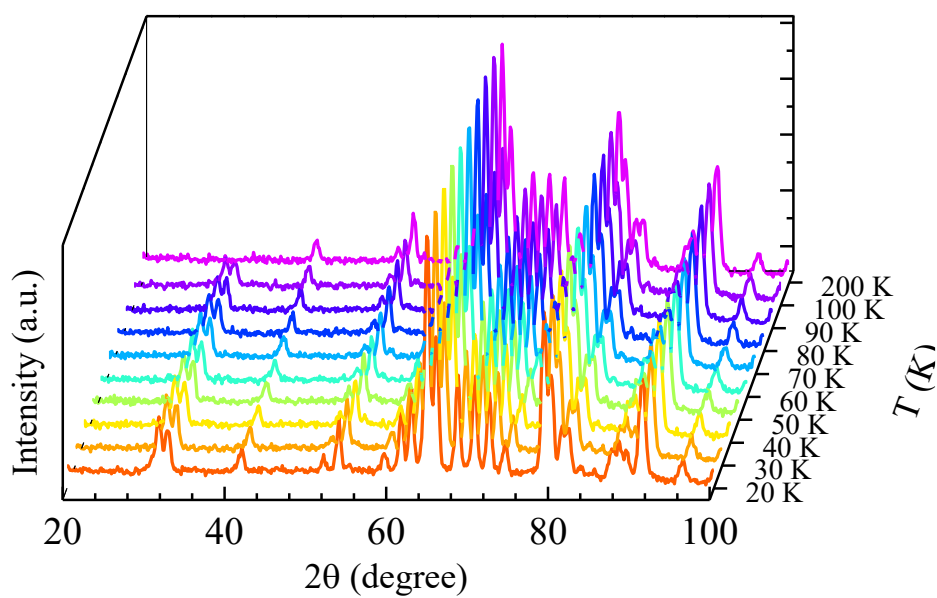

Figure S1: NPD patterns of $TmCrO_3$ collected at different Temperatures.

Fig. S2 presents the refined NPD patterns of collected at 120 K and 50 K, analyzed using the magnetic space group *Pn'ma'* in Γ2 configuration. For the magnetic structure refinement in Γ2 configuration, the symmetry allows all three magnetic moment components of the Cr ions (Ma, Mb, and Mc) to vary freely. However, when all three components are relaxed, the refinements yield nearly negligible values for Ma and Mc. Therefore, additional refinements were performed by constraining Mc = 0, and subsequently by fixing both Ma = Ma = 0. The final refinement is performed by constraining Mc = 0 which does not affect the fit quality, compared to the fully unconstrained model.

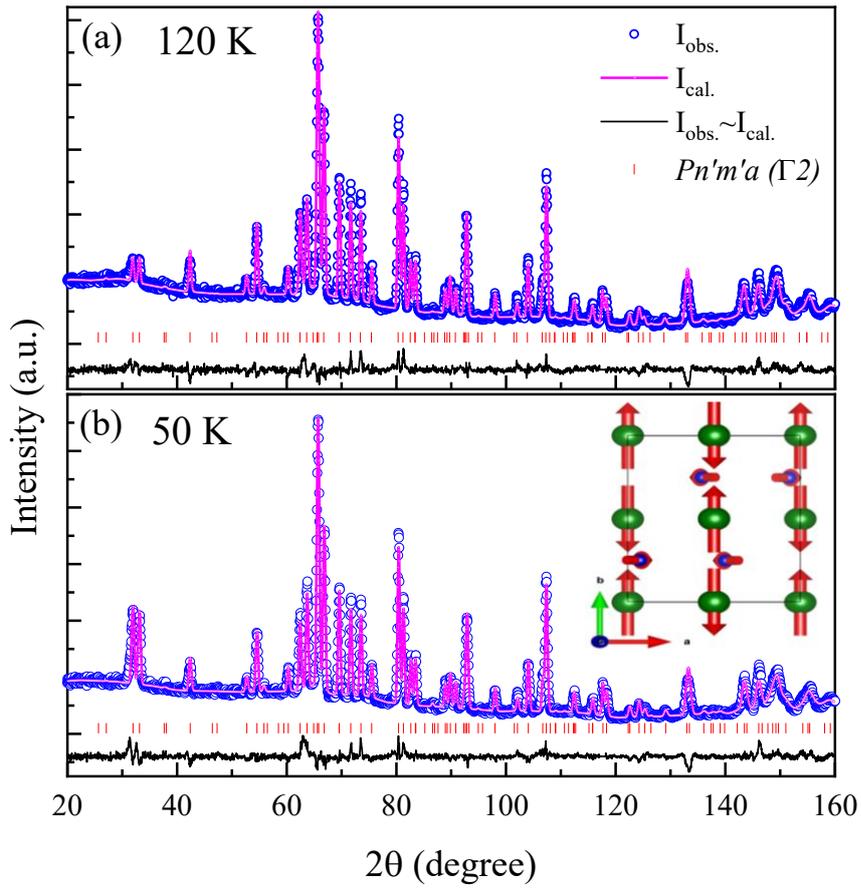

**Figure S2:** Magnetic structure refinement at (a) 120K and (b) 50K using magnetic space group *Pn'm'a* (Γ2). The inset in (b) shows the magnetic structure in Γ2 configuration.